%% file: main.tex
 \documentclass[journal]{IEEEtran}

\normalsize
\usepackage{amssymb,amsmath}
\usepackage{bbding}
\usepackage{flushend,cuted}
\usepackage{amsthm}
\usepackage{amsfonts}
\usepackage{float} 
\usepackage{srcltx}
\usepackage{mathrsfs} 
\usepackage{multirow}
\usepackage{amssymb}
\usepackage{enumitem}

\usepackage{algorithm}
\usepackage[noend]{algpseudocode}
\makeatletter
\def\BState{\State\hskip-\ALG@thistlm}
\makeatother

\usepackage{graphicx}
\usepackage{epsfig}
\usepackage{psfrag}
\usepackage{subfigure}

\usepackage{setspace}
\usepackage{multicol}
\usepackage[noadjust]{cite}
\usepackage{hyperref}

\usepackage[]{units} 
\usepackage{url} 
\usepackage[dvips]{color}
\usepackage{verbatim} 
\usepackage{cite} 
\usepackage{ifthen} 
\usepackage{ifpdf} 
\usepackage{soul}
\usepackage{mathtools}

\newtheorem{definition}{Definition}

\newtheorem{remark}{Remark}

\newcommand\given[1][]{\:#1\vert\:}

\makeatletter
\def\widebreve#1{\mathop{\vbox{\m@th\ialign{##\crcr\noalign{\kern3\p@}%
      \brevefill\crcr\noalign{\kern3\p@\nointerlineskip}%
      $\hfil\displaystyle{#1}\hfil$\crcr}}}\limits}

\def\brevefill{$\m@th \setbox\z@\hbox{$\braceld$}%
  \bracelu\leaders\vrule \@height\ht\z@ \@depth\z@\hfill\braceru$}
\renewcommand*\env@matrix[1][*\c@Ma*matrixCols c]{%
  \hskip -\arraycolsep
  \let\@ifnextchar\new@ifnextchar
  \array{#1}}
\makeatletter

\DeclareRobustCommand{\Nt}[1][Nt]{\ensuremath {N}}
\DeclareRobustCommand{\PrecMat}[1][Nt]{\ensuremath{G}}
\DeclareRobustCommand{\PowerMat}[1][Nt]{\ensuremath{P}}
\ifdefined\BODKIM
\newcommand\figSize{1}
\else
\newcommand\figSize{1}
\fi

\input{eli_macros.tex}

\input{rv_defs.tex}

\begin{document}

\allowdisplaybreaks
%

\title{On the Importance of Asymmetry and Monotonicity Constraints in Maximal Correlation Analysis}

\author{Elad Domanovitz and Uri Erez
}
\maketitle

\begin{abstract}
The  maximal correlation coefficient is a well-established generalization of the Pearson correlation coefficient for  measuring non-linear dependence between random variables. It is appealing from a theoretical standpoint, satisfying R\'enyi's axioms for  measures of dependence. It is also attractive from a computational point of view due to the celebrated alternating conditional expectation algorithm, allowing to compute its empirical version  directly from observed data. Nevertheless, from the outset, it was recognized that the maximal correlation coefficient suffers from some fundamental deficiencies, limiting its usefulness as an indicator of estimation quality.
Another well-known measure of dependence is the  correlation ratio but it too suffers from some drawbacks.
Specifically, the maximal correlation coefficient equals one too easily whereas the correlation ratio equals zero too easily.
The present work recounts some attempts that have been made in the past to alter the definition of the  maximal correlation coefficient in order to overcome its weaknesses and then proceeds to suggest a natural variant of the maximal correlation coefficient.
The proposed dependence measure at the same time resolves the major weakness of the correlation ratio measure and may be viewed as a bridge between the two classical measures.
\end{abstract}

\section{Introduction}
\label{sec:intro}
 Pearson's correlation coefficient is a measure indicating how well one can
approximate (estimate in an average least squares sense) a (response) random variable $Y$ as a linear (more precisely affine) function  of a (predictor/observed) random variable $X$, i.e., as $Y=aX+b$.\footnote{We assume that the random variables $X$ and $Y$ have finite variance.}
The coefficient is given by
\begin{align}
    \rho(X \leftrightarrow Y)=\frac{\rm{Cov}(X,Y)}{\sqrt{\var(X)} \sqrt{ \var(Y)}}.
\end{align}
The coefficient is symmetric in $X$ and $Y$ so it just as well measures how well one can approximate $X$ as a linear function of $Y$.

The correlation ratio of $Y$ on $X$, also suggested by Pearson (see, e.g., \cite{cramer2016mathematical}), similarly measures how well one can approximate $Y$ as a general admissible function of $X$, i.e., as $Y=f(X)$.\footnote{We define a function $f(\cdot)$ to be admissible w.r.t. the random variable $X$ if it is a Borel-measurable real-valued functions such that
$\mathbb{E}[f(X)]=0$ and it has finite and positive variance.}
Specifically, the correlation ratio of $Y$ on $X$ is given by
\begin{align}
    \theta(X \rightarrow Y)
    & =\sqrt{\frac{\var(\mathbb{E}[Y|X])}{\var(Y)}}
    \label{eq:corrRatioEq1}
    \\
    &=\sqrt{1-\frac{\mathbb{E}[\var(Y|X)]}{\var(Y)}}. \nonumber
\end{align}
The correlation ratio can also be expressed as
\begin{align}
    \theta(X \rightarrow  Y)=\sup_{f}\rho(f(X) \leftrightarrow Y)
\label{eq:corr_ratio}
\end{align}
where the supremum is taken over all (admissible) functions $f$ (see, e.g., \cite{renyi1959new}).
This measure is naturally nonsymmetric.
A drawback of the correlation ratio is that it ``equals zero too easily''. Specifically, it can vanish even where the variables are dependent.

We note that one may equivalently say that the correlation ratio measures how well one can approximate
$Y$ as $Y=aX'+b$ for some admissible transformation of the random variable $X'=f(X)$. While perhaps seeming superfluous at this point, this view will prove useful when considering different generalizations of the correlation ratio to the case where the observations are a random vector.


The Hirschfeld-Gebelein-R\'enyi maximal correlation coefficient \cite{hirschfeld1935connection,gebelein1941statistische,renyi1959measures} measures the maximal (Pearson) correlation that can be attained by transforming the pair $X,Y$ into random variables $X'=g(X)$ and $Y'=f(Y)$; that is, how well $X'=aY'+b$ holds in a mean squared error sense for some pair of functions $f$ and $g$. More precisely,  the maximum correlation coefficient is defined as the supremum over all (admissible) functions $f,g$ of the correlation between $f(X)$ and $g(Y)$:
\begin{align}
   \rho^{**}_{\rm max}(X \leftrightarrow Y)=\sup_{f,g}\rho(f(X) \leftrightarrow g(Y)).
\end{align}
This measure is again symmetric by definition. We use the superscript  ``**'' to indicate that both functions (applied to the response and the predictor random variables) need not satisfy any restrictions beyond being admissible.

The maximal correlation coefficient has some very pleasing properties. 
In particular, in \cite{renyi1959measures}, R\'enyi 
put forth a set of seven axioms deemed natural to require of a  measure of dependence between
a pair of random variables and established that the maximal correlation coefficient satisfies the full set of axioms. In particular, unlike the correlation ratio, the maximal correlation coefficient ``does not equal zero too easily''. Further, unlike the correlation ratio it is symmetric, which was set as one of the axioms. Nonetheless, these pleasing properties come at the price of introducing new problems as elaborated on below.



Another appealing trait of the maximal correlation coefficient, greatly contributing to its popularity, is its tight relation to 
a Euclidean geometric framework and to  operator theory. In particular, it is readily computable numerically via the alternating conditional expectation (ACE) algorithm of Breiman and Friedman \cite{breiman1985estimating}.
Moreover, and as recalled in the sequel, the ACE algorithm naturally extends to cover linear estimation of a (transformed) random variable from a component-wise transformed random vector.

R\'enyi's seminal work inspired substantial subsequent work aiming to identify other measures of dependence satisfying the set of axioms. We refer the reader to \cite{samuel2001correlation} for a survey of some of these.

Despite its elegance and it being amenable to computation, the maximal correlation coefficient suffers from some significant deficiencies as was recognized  since its inception.
Specifically, it is well known that it ``equals one too easily''; see, e.g. \cite{hall1967characterizing}, \cite{kimeldorf1978monotone}, as well as Footnote~3 in \cite{renyi1959measures}.
In fact, 
it can equal one even for  two random variables that are nearly independent (as also demonstrated below).

Disconcerted by this behavior of the maximal correlation coefficient, Kimeldorf and Sampson \cite{kimeldorf1978monotone}
proposed to alter its definition, 
introducing monotonicity constraints. They defined a monotone dependence measure as follows. 
\vspace{-0mm}
\begin{align}
    \rho^{mm}_{\rm max}(X \leftrightarrow Y)=\sup_{f,g}\rho(f(X) \leftrightarrow g(Y)),
    \label{eq:MonotoneDependence}
\end{align}
where $f$ and $g$ are not only admissible but also monotone. 
Nevertheless, as stated in \cite{kimeldorf1978monotone}, while the imposed constraints somewhat mitigate the ``easiness of attaining the value of one", the measure \eqref{eq:MonotoneDependence} still can equal one for a pair of random variables that are not completely dependent.  


The definition of the monotone dependence measure \eqref{eq:MonotoneDependence} is unsatisfactory in two respects. The first is that is imposes 
symmetric constraints on the two transformations. As the process of estimation/prediction (and more generally inference) is directional, if  the goal of the dependence measure is to characterize how well one can achieve the latter tasks, there is no apparent reason to impose any restriction on the transformation applied to the observed data. In this respect, 
it is worth quoting the incisive comments (in reference to \cite{ramsay1988monotone}) of  Hastie and Tibshirani   \cite{hastie1988monotone}:

\emph{``\ldots a monotone restriction makes sense for a response transformation because it is necessary to allow predictions of the response from the estimated model. \ldots~On the other hand, why restrict
 predictor transformations (such as for displacement and weight in the city gas consumption problem) to be monotone? Instead, why not leave them unrestricted and let the data suggest the shape of the relevant transformation?"}

The second and more subtle deficiency of  the monotone dependence measure of Kimeldorf and Sampson (as well as the semi-monotone variant suggested by Hastie and Tibshirani) is that when it comes to the response variable, the requirement that the transformation be  monotone is not strong enough. 

The goal of the present work is first to reiterate some of the known drawbacks of both the correlation ratio and of the maximal correlation coefficient,  and then to suggest a possible resolution. 
In particular, we demonstrate
that while allowing  a transformation to be applied to the response variable is important, it is not sufficient to require that it be monotonic.
Rather, one must strengthen the required ``degree'' of monotonicity. 

Specifically, we introduce the notion of $\kappa$-monotonicity and  argue in favor of constraining (only) the transformation applied to the response random variable to be $\kappa$-monotonic, leading to a proposed semi-$\kappa$-monotone maximal correlation measure. 
The parameter $\kappa$ dictates a minimal and maximal slope that the function applied to the response variable must maintain.

We show that requiring that $0<\kappa<1$ yields a  measure that does not suffer from  the drawbacks of neither the maximal correlation coefficient nor from those of the correlation ratio. 
The  proposed measure satisfies a set of modified R\'enyi axioms that does not sacrifice the natural requirements  of capturing {\em both} independence and complete dependence. 

It is important to note that setting $\kappa=0$ amounts to requiring merely monotonicity,  as already suggested in \cite{hastie1988monotone}, without imposing a minimal (or maximal) positive slope. Setting $\kappa=1$, on the other hand, the  semi-$\kappa$-monotone maximal correlation measure  reduces to the correlation ratio. This implies that while both of these choices are not sufficient to satisfy the proposed set of modified R\'enyi axioms, they lie right on the boundary of the set of values that do. Thus, in a sense, these choices may be considered satisfactory if we rule out ``pathological'' examples. The same \emph{cannot} be said of the maximal correlation coefficient. These points are illustrated in Section~\ref{sec:numericalExample}.

Finally, as the usefulness of the maximal correlation coefficient is due in part to it being readily computable, we suggest modifications to the ACE algorithm and exemplify the resulting performance via several examples.

The rest of the paper is organized as follows. Section~\ref{sec:short} describes the shortcomings of the correlation ratio and maximal correlation coefficient as well as presents the proposed dependence measure. Section~\ref{sec:mod_ren}
puts forth a set of modified R\'enyi axioms, outlining the main steps involved in proving that the proposed measure satisfies the latter. Section~\ref{sec:modifiedAce} provides some modifications to the ACE algorithm, enforcing monotonicity constraints. Finally, several numerical examples illustaring the main ideas are given in Section~\ref{sec:numericalExample}.

\section{Shortcomings of the Correlation Ratio and Maximal  Correlation Coefficient and a Proposed Resolution}
\label{sec:short}

As a simple example, consider two (sequences of) random variables that share only the least significant bit:
\begin{align}
X^{(N)} &=C +\sum_{i=1}^N A_i 2^i \nonumber  \\
Y^{(N)} &=C +\sum_{i=1}^N B_i 2^i,
\label{eq:example0}
\end{align}
where ${A_i},{B_i},C$ are mutually independent random variables, all taking the values $0$ or $1$ with equal probability.
Clearly, applying  modulo $2$ to both variables yields
a correlation of one.
This seems quite unsatisfactory if our goal is estimation subject to any reasonable distortion metric as the two random variables become virtually independent as $N$ grows. Specifically, the pair $(X^{(N)}/2^N,Y^{(N)}/2^N)$ converges in distribution to a uniform distribution over the unit square.
\begin{remark}
It should be noted in this respect that the maximal correlation coefficient is a good measure for a different goal. It quantifies to what extent two random variables share any common ``features". 
\end{remark}

\begin{remark}[Discrete random variables]
While the emphasis in this paper is on continuous random variables, symmetric measures are also generally not appropriate for measuring the dependence between discrete random variables.
For instance, a natural measure in this case is the minimal possible probability of error when predicting one from the other.
Clearly, this measure is also not symmetric.
While minimum error probability is related via  universal lower and upper bounds to the conditional entropy and mutual information (the latter being a symmetric measure), as shown in \cite{tebbe1968uncertainty} (Equations 5 and 6), the gap between the lower and upper bounds (keeping the probability of error fixed) grows unbounded with the cardinality of the random variables.
This is yet another indication that symmetric measures are ill-suited for estimation/prediction purposes.
\end{remark}

A natural and quite satisfying  measure of directional dependence between random variables, that takes the value of one only when the response variable is a function of the predictor variable, is the correlation ratio defined in \eqref{eq:corrRatioEq1}.
While R\'enyi objected to the correlation ratio due to  its asymmetric nature, as was noted in \cite{hall1967characterizing}, when our goal is asymmetric  (i.e., estimating $Y$ from $X$), there is no reason for requiring that the measure be symmetric.

Nonetheless, in some cases one does not have strong grounds to assume a particular ``parameterization" of the desired (response) random variable. Thus, not allowing to apply any transformation to the response variable, as is the case of the correlation ratio, may  be too restrictive.
In other words, in the absence of a preferred ``natural" parameterization of the response variable, one may consider choosing a strictly monotone transformation (change of variables) so as to make it  easier to estimate.
A more severe drawback of the correlation ratio is that it vanishes too easily, i.e., it can be zero for two dependent random variables. 

In light of these considerations, we propose the following modification to the definition of the maximal correlation coefficient.

\begin{definition}
\label{def:eps_increasing}
For $0 \leq \kappa \leq 1$, a function $f$ is said to be $\kappa$-increasing, if for all $x_2 \geq x_1$:
\begin{align*}
    f(x_2)-f(x_1) & \geq \kappa(x_2-x_1) \, , \nonumber \\
    f(x_2)-f(x_1) & \leq \frac{1}{\kappa}(x_2-x_1).
\end{align*}
\vspace{-1mm}
\end{definition}

\begin{definition}
 \label{def:semiMonotoneMeasure}
For a given $0<\kappa<1$, the semi-$\kappa$-monotone maximal correlation measure is defined as
\begin{align*}
    \rho^{*m_{\kappa}}_{\rm max}(X\rightarrow Y)=\sup_{f,g}\rho(f(X)\leftrightarrow g(Y) )
\end{align*}
where and the supremum is taken over all admissible functions $f$, and over  $\kappa$-increasing admissible functions $g$.
\end{definition}

\begin{remark}
Limiting $g$ to be $\kappa$-increasing implies that, in particular, it is invertible, which is a natural requirement. Moreover, the set of $\kappa$-increasing admissible functions is closed.
We further note that the value of $\kappa$ controls how far the measure can deviate from the correlation ratio.
\end{remark}





\subsection{The vector observation case}

Let   $\svv{X}=(X_1,\ldots,X_p)$ be a vector of predictor variables.
The maximal correlation coefficient becomes
 \begin{align}
     \rho^{**}_{\rm max}( \svv{X} \leftrightarrow Y)=\sup_{f,g}\rho(f(\svv{X}) \leftrightarrow g(Y))
 \end{align}
where the supremum is over all admissible functions.

Following Breiman and Friedman \cite{breiman1985estimating}, we may also consider a simplified (quasi-additive) relationship between $Y$ and $\svv{X}$
where we seek an optimal linear regression between a transformation of $Y$ and a component-wise non-linear transformation of the predictor random vector $\svv{X}$. Denote the fraction of the variance not explained by a regression of $f(Y)$ on $\sum_i f_i(X_i)$ as
\begin{align}
    e^2(g,f_1,\ldots,f_p)=\frac{\mathbb{E}\left[\left(g(Y)-\sum_i f_i(X_i)\right)^2\right]}{\mathbb{E} [g(Y)^2]}
\end{align}
where zero-mean functions are assumed.
In \cite{breiman1985estimating}, conditions for the existence of optimal transformations $\{f_i\},g$ such that the supremum is attained are given, and it is shown that under these conditions the ACE algorithm converges to the optimal transformations.

Going back to the rationale for requiring $\kappa$-monotonicity, one may object to the example \eqref{eq:example0} as being artificial and argue that the maximum correlation coefficient merely captures whatever dependence there is between the random variables. In this respect, it is worthwhile quoting Breiman \cite{breiman1988monotone} (commenting on \cite{ramsay1988monotone}):

\emph{``I only know of infrequent cases in which I would
 insist on monotone transformations. Finding non-monotonicity can lead to interesting scientific discoveries. If the appropriate transformation is monotone,
 then the fitted spline functions (or ACE transformations) will produce close to a monotonic transformation. So it is hard to see what there is to gain in the
 imposition of monotonicity.''}

We now demonstrate  that the problematic nature of the maximal correlation coefficient becomes more pronounced when considering the multi-variate case and so does the necessity of restricting the transformation of the response variable (only) to be 
monotone.

Specifically, let us consider again the example of \eqref{eq:example0}.
Suppose that $Y$ and $X$ are as defined but that in addition to $X$, there is another slightly noisy observation of $Y$, say $\tilde{X}=Y+Z$ where the variance of $Z$ is small with respect to that of $Y$. Clearly, the maximal correlation coefficient will still equal $1$, and the observation $\tilde{X}$ will be discarded even though it could have allowed to estimate $Y$ with small distortion.
Thus, in this example, the maximal correlation coefficient is maximized by perfectly estimating the least significant bit while doing away with the more significant bits even though nearly distortionless reconstruction is possible.
See also Section~\ref{sec:numericalExample} below for a numerical
example.

\section{modified R\'enyi axioms}
\label{sec:mod_ren}

We follow the approach of Hall \cite{hall1967characterizing} in defining an asymmetric variant of the  R\'enyi axioms; more precisely, we adopt a slight variation on the somewhat stronger version formulated by Li \cite{li2016true}. However, unlike both of these works, when it comes to  putting forward a candidate dependence measure satisfying the modified axioms, we adhere to a mean square error methodology.


Assume $r(X\rightarrow Y)$  is to measure the degree of dependence of $Y$ on $X$. Then we require that it satisfy the following:
\begin{enumerate}[label=(\alph*)]
\item \mbox{$r(X\rightarrow Y)$} is defined for all non-constant  random variables $X,Y$ having finite variance.\footnote{In \cite{li2016true}, the first axiom only requires that $r(X\rightarrow Y)$ be defined  for  \emph{continuous} random variables $X,Y$.
}
\item $r(X\rightarrow Y)$ may not be equal to $r(Y\rightarrow X)$.
\item $0\leq r(X\rightarrow Y)\leq 1$.
\item $r(X\rightarrow Y)=0$ if and only if $X,Y$ are independent.
\item $r(X\rightarrow Y)=1$ if and only if $Y=f(X)$ almost surely for some admissible function $f$.
\item If $f$ is an admissible bijection on $\mathbb{R}$, then $r(f(X)\rightarrow Y)=r(X\rightarrow Y)$
\item If $X,Y$ are jointly normal with correlation coefficient $\rho$, then $r(X\rightarrow Y)=|\rho|$.\footnote{In \cite{li2016true}, the last axiom only requires that if $X,Y$ are jointly normal with correlation coefficient $\rho$, $r(X\rightarrow Y)$ is a strictly
increasing function of $|\rho|$.}
\end{enumerate}

\begin{remark}
We note that the correlation ratio, defined in \eqref{eq:corr_ratio},  satisfies all of the modified axioms except for the ``only if'' part of axiom (d).
\end{remark}


We next observe that for  absolutely continuous (or discrete) distributions, the
semi-$\kappa$-monotone maximal correlation measure of Definition~\ref{def:semiMonotoneMeasure} 
satisfies the proposed axioms.

It is  readily verified that axioms (a), (b) and (c) hold.
To show that axiom (d) holds, we note that if $X,Y$ are independent, then obviously \mbox{$\rho^{*m_{\kappa}}_{\rm max}(X\rightarrow Y)=0$}, as so is even \mbox{$\rho^{**}_{\rm max}(X\leftrightarrow Y)$}. As for the other direction, we first note that it suffices to consider the case where the correlation ratio equals $0$ and $X,Y$ are dependent. Since the correlation ratio is $0$, it follows
from \eqref{eq:corrRatioEq1} that $\mathbb{E}[Y|X] \equiv const$ (in the mean square sense), i.e., 
\begin{align}
    \int p(y|x)y dy={\rm const}.
\end{align}
We may break the symmetry of $g(y)=y$ by defining, e.g.,
 \begin{align}
g_{a,\kappa}(y)=\begin{cases}
y~~~y\geq a \\
\kappa y~~y<a
\end{cases}.
\label{eq:g_a_k}
 \end{align}
Consider two values of $x_1$ and $x_2$ for which
 the functions $p(y|x_i)$ are not identical (as functions of $y$), as must exist by the assumption of dependence.
Let $a$  be a value such that
 \begin{align}
     \int^a p(y|x_1)ydy \neq  \int^a p(y|x_2)ydy.
 \end{align}
Without loss of generality, we may assume that the left hand side is smaller than the right hand side (we may rename $x_1$ and $x_2$).
Recalling that $\kappa<1$, it follows that
 \begin{align}
     \int p(y|x_1)g_a(y)dy > \int p(y|x_2)g_a(y)dy
 \end{align}
Thus,
$$\mathbb{E}[g_a(Y)|X=x_1] \neq \mathbb{E}[g_a(Y)|X=x_2]$$
and hence the correlation ratio of $Y'=g_a(Y)$ on $X$ is non-zero, giving a lower bound to the semi-$\kappa$-monotone maximal correlation measure between $Y$ and $X$.  Note that~\eqref{eq:g_a_k} imposes only a lower bound on the slope of the function applied to the response variable.



To show that axiom (e) holds, we note that by definition, if $Y=f(X)$ (almost surely), then $\rho^{*m_{\kappa}}_{\rm max}(X\rightarrow Y)=1$. To show that the opposite direction holds, we first note that it can be shown that the supremum in  \eqref{def:semiMonotoneMeasure} is attained. 
Recalling that if $\rho^{*m_{\kappa}}_{\rm max}(X\rightarrow Y)=1$, then by the properties of Pearson's correlation coefficient, there is a \emph{perfect} linear regression between $g(Y)$ and $f'(X)$ ($g,f'$ being maximizing functions of the measure).
Now, it can be shown that the supremum 
Hence, we have $g(Y)=af'(X)+b$ where $g$ is an increasing function with slope greater than $\kappa$. 
Since $\kappa$ is strictly positive, it follows that not only is $g$ invertible, but also $g^{-1}(Y)$ has finite variance (since the slope of $g^{-1}(Y)$ is at most $\frac{1}{\kappa}$ and $Y$ has finite variance). Therefore we have $Y=g^{-1}(af'(X)+b)$. Denoting $f(X)=g^{-1}(af'(X)+b)$, we note that if $f'$ is admissible, then so is $f$.

Axiom (f) trivially holds.
To show that axiom (g) holds, we recall that it is well known that when $X,Y$ are jointly normal with correlation coefficient $\rho$, then \mbox{$\rho^{**}_{\rm max}(X\leftrightarrow Y)=|\rho|$} (see, e.g., \cite{lancaster1957some} and \cite{yu2008maximal}). Since this implies that the maximal correlation is achieved taking $g(y)=y$ (i.e., a monotone function) and $f(x)=x$ or $f(x)=-x$, it follows that
\begin{align}
    \rho^{*m_{\kappa}}_{\rm max}(X\rightarrow Y)&=\rho^{**}_{\rm max}(X\leftrightarrow Y) \nonumber \\
&=|\rho|.
\end{align}


We note that the restriction $0 < \kappa < 1$ is necessary. Specifically, for $\kappa=1$, axiom (d) is not satisfied whereas for $\kappa=0$,  axiom (e) is not satisfied.

Finally, we note that one may define other dependence measures satisfying   modified R\'enyi axioms, most notably via the theory of copulas (which is inherently related to monotone constraints); e.g., a symmetric measure is given in  \cite{schweizer1981nonparametric} and a directional one is given in  \cite{li2016true}.
Nonetheless, we believe that the proposed
measure has the advantage of being closely tied to linear regression methods and geometric considerations.

\section{Modified ACE Algorithm}
\label{sec:modifiedAce}
We begin by presenting a modification of the ACE algorithm  with the goal of computing 
the semi-$0$-monotone maximal correlation measure  $\rho^{*m_{0}}_{\rm max}(X\rightarrow Y)$,  
restricting the function applied to the response variable only to be weakly monotone.


As we do not know of a simple means to enforce the slope constraints, we do not have an algorithm for computing the semi-$\kappa$-monotone maximal correlation measure. Instead, we have employed a regularized version of the original ACE algorithm as described in Section~\ref{sec:reglar}.

\subsection{Evaluating the semi-$0$-monotone maximal correlation measure}
\label{sec:modACE}

We now present a modification of the ACE algorithm to compute 
the semi-$0$-monotone maximal correlation measure $\rho^{*m_{0}}_{\rm max}(X\rightarrow Y)$ for the case of single predictor variable. It is readily seen that the correlation increases in each iteration of the algorithm and thus converges but we do not pursue proving optimality.
We then generalize the algorithm to the quasi-additive multi-variate scenario.




Following in the footsteps of
\cite{breiman1985estimating}, recall that the space of all random variables with finite variance is a Hilbert space, which we denote by $\mathcal{H}_2$, with the usual definition of the inner product \mbox{$<X,Y>=\mathbb{E}[X Y]$}, for $X,Y\in\mathcal{H}_2$.
We may further define the subspace $\mathcal{H}_2(X)$ as the set of all random variables that correspond to an admissible function of $X$. We similarly define the subspace $\mathcal{H}_2(Y)$.
Now, if we further limit the functions applied to $Y$ to be non-decreasing, 
we obtain a closed and convex subset of the Hilbert space $\mathcal{H}_2(Y)$. We denote this set by $\mathcal{M}_{0}(Y)$.

Denoting by $P_{\mathcal{A}}(Y)$  the orthogonal projection of $Y$ onto the closed convex set $\mathcal{A}$,\footnote{Note that $\mathcal{P}_{\mathcal{H}_2(X)}\left(g(Y)\right)=\mathbb{E}\left[ g(Y)\given X \right]$.} the modified ACE algorithm is described in Algorithm~\ref{Alg:ACE_Semi_0_mono_Single_Predictor} for the case of single predictor variable.
\begin{algorithm}
\caption{}\label{Alg:ACE_Semi_0_mono_Single_Predictor}
\begin{algorithmic}[1]
\Procedure{Calculate-Semi-$0$-monotone}{}
\State Set $g(Y)=Y/\|Y\|$;
\While{$e^2(g,f)$ decreases}
            \State $f'(X)=\mathcal{P}_{\mathcal{H}_2(X)}\left(g(Y)\right)$
             \State replace $f(X)$ with $f'(X)$
    \State $g'(Y)=\mathcal{P}_{\mathcal{M}_{0}(Y)}\left(f(X)\right)$
    \State replace $g(Y)$ with $g'(Y)/\|g'(Y)\|$
\EndWhile
\State End modified ACE
\EndProcedure
\end{algorithmic}
\end{algorithm}

In the case of a multi-variate predictor, the original ACE algorithm seeks an optimal linear regression between a transformation of $Y$ and a component-wise non-linear transformation of the predictor random vector $\bf{X}$. The latter transformations are defined by a set of admissible functions $f_1,\ldots,f_p$, each function operating on the corresponding random variable, yielding an estimator of the form $\sum_i f_i(X_i)$.

Therefore Algorithm~\ref{Alg:ACE_Semi_0_mono_Single_Predictor} becomes
\begin{algorithm}
\caption{}\label{Alg:ACE_Semi_0_mono_Multi_Predictor}
\begin{algorithmic}[1]
\Procedure{Calculate-Semi-$0$-monotone}{}
\State Set $g(Y)=Y/\|Y\|$ and $f_1(x_1),\cdots,f_p(x_p)=0$;
\While{$e^2(g,f_1,\ldots,f_p)$ decreases}
    \While{$e^2(g,f_1,\ldots,f_p)$ decreases}
        \For{\texttt{$k=1$ to $p$}}
            \State $f'_{k}(X_k)=$ \newline
            \hspace*{6em} $\mathcal{P}_{\mathcal{H}_2(X_k)}\left(g(Y)-\sum_{i\neq k}f_{i}(X_i)\right)$
            \State replace $f_k(X_k)$ with $f'_{k}(X_k)$ 
        \EndFor
    \EndWhile
    \State $g'(Y)=\mathcal{P}_{\mathcal{M_{0}}(Y)}(Y)\left(\sum_i f_{i}(X_i) \right)$
    \State replace $g(Y)$ with $g'(Y)/\|g'(Y)\|$
\EndWhile
\State End modified ACE
\EndProcedure
\end{algorithmic}
\end{algorithm}

\subsection{Regularized ACE algorithm}
\label{sec:reglar}
We may enforce that the transformation $g$ be
$\kappa$-monotone by applying the  following regularization. Given a monotone transformation $g$ (e.g., the outcome of Algorithm~1), do:\footnote{Note that this method of regularization actually forces the slope to be between $\kappa$ and  $1/\kappa+\kappa$.
}
\begin{align*}
        g'(Y)&=g^{-1}(Y)+\kappa\cdot Y \\
        g(Y)&=g'^{-1}(Y)+\kappa\cdot Y
\end{align*}

\begin{algorithm}
\caption{}\label{Alg:ACE_kappa_inc_fixed_norm}
\begin{algorithmic}[1]
\Procedure{Regularized-ACE}{}
\State Set $g(Y)=Y/\|Y\|$;
\While{$e^2(g,f)$ decreases}
            \State $f'(X)=\mathcal{P}_{\mathcal{H}_2(X)}\left(g(Y)\right)$
             \State replace $f(X)$ with $f'(X)$
    \State $g'(Y)=\mathcal{P}_{\mathcal{M}_{0}(Y)}\left(f(X)\right)$
    \State replace $g(Y)$ with $g'(Y)/\|g'(Y)\|$
\EndWhile
\State Apply regularization
\State End regularized ACE
\EndProcedure
\end{algorithmic}
\end{algorithm}
A similar regularization can be applied  to Algorithm~\ref{Alg:ACE_Semi_0_mono_Multi_Predictor}.
\section{Numerical examples}
\label{sec:numericalExample}

As we do not know of an efficient means to evaluate the
semi-$\kappa$-monotone maximal correlation measure for $\kappa\neq 0$, we for the most part demonstrate the advantage of the semi-$0$-monotone maximal correlation measure over the standard maximal correlation measure, 
in the context of estimation of a random variable $Y$ from a random vector $\bf{X}$. We further demonstrate its potential for improvement over the correlation ratio. To that end, we compute the semi-$0$-monotone maximal correlation measure using Algorithm 1 presented in Section~\ref{sec:modifiedAce}.


We begin with a multi-variate example where one of the two observed random variables ``masks'' the other while the latter is more significant for estimation purposes. We then demonstrate why taking $\kappa=1$ is inadequate and hence the semi-$\kappa$-monotone maximal correlation measure may be advantageous with respect to the correlation ratio.
The third example illustrates why taking $\kappa=0$ is not sufficient in general, and  heuristically demonstrates that Algorithm~\ref{Alg:ACE_kappa_inc_fixed_norm} yields more satisfying results.

For simulating ACE, we used the ACE Matlab code provided by the authors of \cite{voss1997reconstruction}. 
To limit $g$ to be a monotonic function, we used isotonic regression.

\subsection{Example 1 - Multi-variate predictor}

Assume that the response variable $Y$ is distributed uniformly over the interval $[0,1]$ and that we have two predictor variables
\begin{align}
    X_1&=\rm{mod}(Y,0.2)+N_1 \nonumber \\
    X_2&=Y^3+N_2,
    \label{eq:example1}
\end{align}
where $N_1,N_2$ are independent zero-mean Gaussian variables with  $\sigma^2_{N_1}=0.01$ and $\sigma^2_{N_2}=0.2$.

\begin{figure}[htbp]
    \centering
    \includegraphics[width=\figSize\columnwidth]{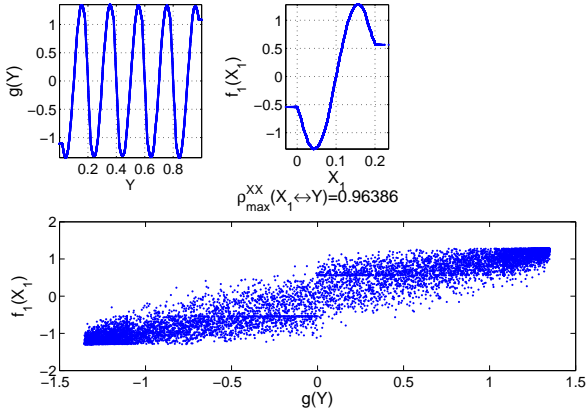}
    \caption{Example 1: Running ACE on $Y$ and $X_1$.}
    \label{fig:y_x1_ACE}
\end{figure}

The optimization invloved in the maximal correlation coefficient results in ``shadowing'' the more significant variable, $X_2$, for estimation purposes of the response $Y$. To see this, we start by running the ACE algorithm to evaluate the maximal correlation coefficient between $Y$ and $X_1$. As can be seen from Figure~\ref{fig:y_x1_ACE}, this results in a very high value. Inspecting the transformations yielding this result, we observe that $g$ is not monotonic and hence we cannot recover $Y$ from $g(Y)$. 

\begin{figure}[htbp]
    \centering
    \includegraphics[width=\figSize\columnwidth]{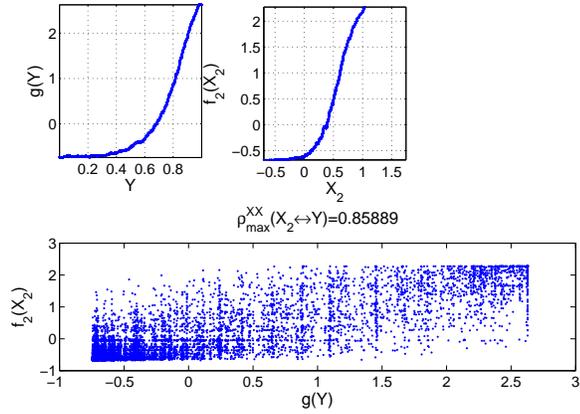}
    \caption{Example 1: Running ACE on $Y$ and $X_2$.}
    \label{fig:y_x2_ACE}
\end{figure}

Next, we apply the ACE algorithm to calculate the maximal correlation coefficient between $Y$ and $X_2$. As can be seen from Figure~\ref{fig:y_x2_ACE},  this value is much smaller (than that between $Y$ and $X_1$) since in this case we have stronger additive noise. Nevertheless, the transformation applied to $Y$ is now monotonic. Therefore, even though the maximal correlation coefficient is smaller, the observation $X_2$ can better serve the purpose of estimation of $Y$.

\begin{figure}[htbp]
    \centering
    \includegraphics[width=\figSize\columnwidth]{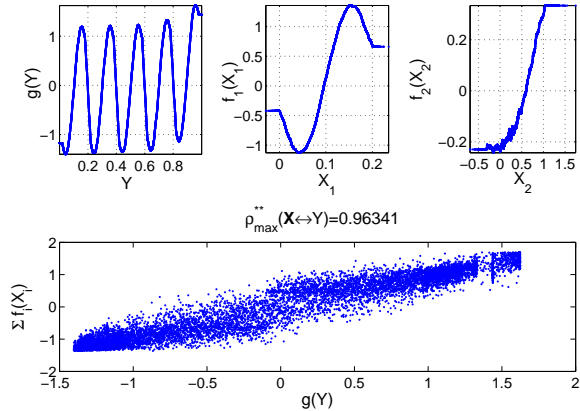}
    \caption{Example 1: Running ACE on $Y$, $X_1$ and $X_2$.}
    \label{fig:y_x1_x2_ACE}
\end{figure}

Next we apply the ACE algorithm to 
$Y$ and the
vector $(X_1,X_2)$. As can be seen from Figure~\ref{fig:y_x1_x2_ACE}, the ACE algorithm, in order to maximize the  correlation, chooses similar functions as in case of running only on $Y$ and $X_1$,  practically choosing to ignore $X_2$. While, indeed, this maximizes the correlation coefficient, it is far from satisfying from an estimation viewpoint.

\begin{figure}[htbp]
    \centering
    \includegraphics[width=\figSize\columnwidth]{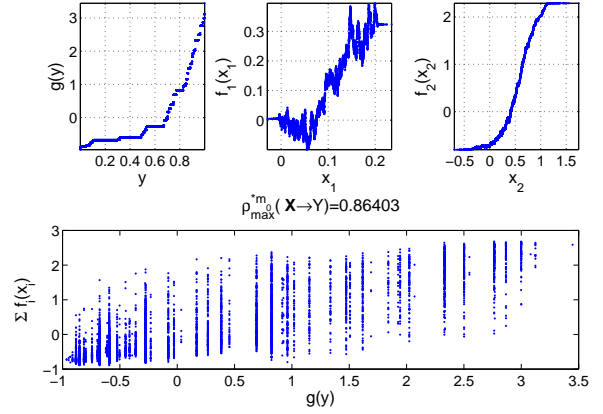}
    \caption{Example 1: Running modified ACE (Algorithm~\ref{Alg:ACE_Semi_0_mono_Multi_Predictor}) on $Y$, $X_1$ and $X_2$ with $\kappa=0$.}
    \label{fig:y_x1_x2_SemiMon_stric}
\end{figure}

As can be seen from Figure~\ref{fig:y_x1_x2_SemiMon_stric},
the resulting value of the semi-$0$-monotone maximal correlation measure is very close to the maximal correlation value between $Y$ and $X_2$. Thus, the algorithm ``chooses to ignore'' $X_1$ (even though it suffers from a lower noise level) and bases the estimation on $X_2$.



\subsection{Comparisons with the correlation ratio}
\subsubsection{Example 2a}
The proposed dependence measure can be viewed as a generalization of the correlation ratio (the correlation ratio amounts to setting $g$ to have a constant slope of $1$).

\begin{figure}[htbp]
    \centering
    \includegraphics[width=\figSize\columnwidth]{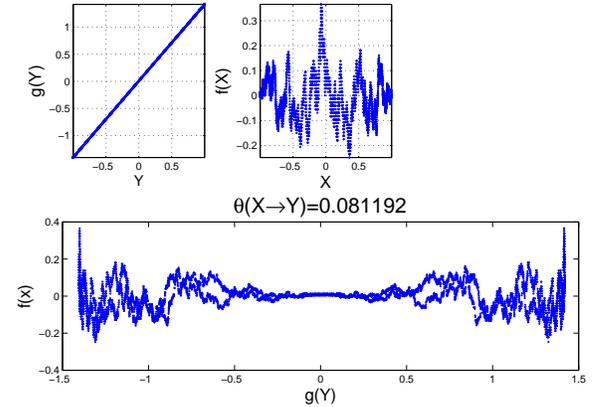}.
    \caption{Example 2a: Optimal transformation corresponding to the correlation ratio.}
    \label{fig:y_x_exmp4_corr_ratio}
\end{figure}

We first demonstrate how the semi-$\kappa$-monotone maximal correlation measure deals with a well-known example where the correlation ratio equals $0$ for a pair of dependent random variables. Specifically, we consider a vector $(X,Y)$  that is uniformly distributed over a circle with radius $1$.  The correlation ratio is $0$ as depicted in Figure~\ref{fig:y_x_exmp4_corr_ratio} where we ran ACE enforcing $g(y)=y$.

\begin{figure}[htbp]
    \centering
    \includegraphics[width=\figSize\columnwidth]{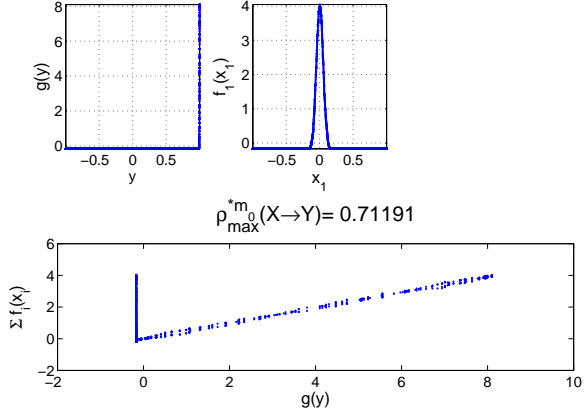}.
    \caption{Example 2a: Running modified ACE (Algorithm~\ref{Alg:ACE_Semi_0_mono_Single_Predictor}) on $Y$ and $X_1$ with $\kappa=0$.}
    \label{fig:y_x_exmp4_semi_mono_0_1}
\end{figure}

Applying Algorithm~\ref{Alg:ACE_Semi_0_mono_Single_Predictor} with $\kappa=0$ yields a much larger correlation. Thus, it manages to capture the dependence between $X$ and $Y$. This is depicted in Figure~\ref{fig:y_x_exmp4_semi_mono_0_1}.

\subsubsection{Example 2b}
The next example demonstrates another potential advantage over the correlation ratio.
As was already noted, there are cases where there is no a priori preferred (natural) parameterization for the
response variable and thus choosing one that maximizes the correlation may be a reasonable approach as we now demonstrate.

\begin{figure}[htbp]
    \centering
    \includegraphics[width=\figSize\columnwidth]{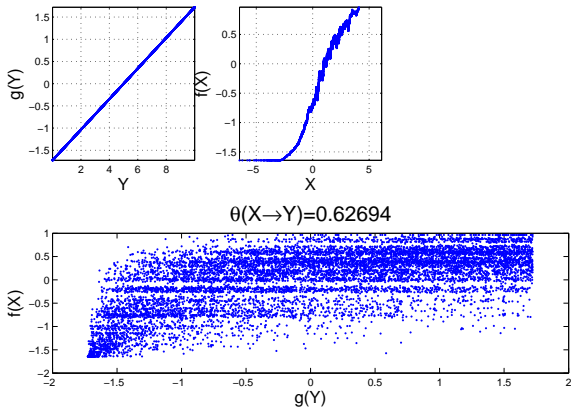}.
    \caption{Example 2b: Optimal transformations corresponding to the correlation ratio.}
    \label{fig:y_x_exmp3_Corr_Ratio}
\end{figure}

\begin{figure}[htbp]
    \centering
    \includegraphics[width=\figSize\columnwidth]{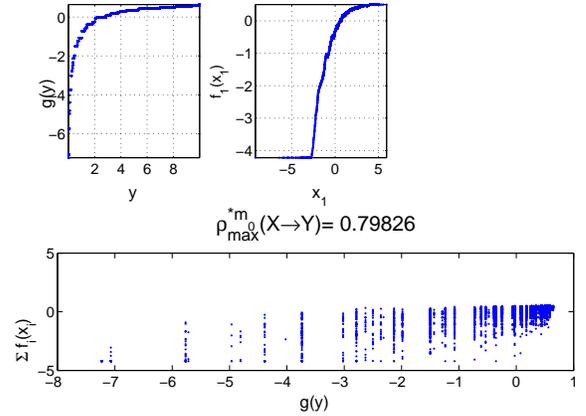}
    \caption{Example 2b: Running modified ACE (Algorithm~\ref{Alg:ACE_Semi_0_mono_Single_Predictor}) on $Y$ and $X_1$ with $\kappa=0$.}
    \label{fig:y_x_exmp3_semi_0_1}
\end{figure}

Assume that the response variable $Y$ is distributed uniformly over the interval $[0,10]$ and that the predictor variable $X$ is
\begin{align}
    X=\log(Y)+N,
    \label{eq:example3}
\end{align}
where $N$ is a zero-mean Gaussian (and independent of $Y$) with unit variance. 
Comparing the correlation ratio (Figure~\ref{fig:y_x_exmp3_Corr_Ratio}) and the results of Algorithm~\ref{Alg:ACE_Semi_0_mono_Single_Predictor} with $\kappa=0$ (Figure~\ref{fig:y_x_exmp3_semi_0_1}), reveals that the correlation of the latter is significantly higher.

\subsection{Example 3 - Semi-$0$-monotonicity is insufficient}

To illustrate why it does not suffice to limit $g$ to be merely monotone, consider the following example. Assume that the response $Y$ is distributed uniformly over the interval $[-10,10]$ and that
\begin{align}
    X=\begin{cases}
    X=Y~~~Y>9 \\
    X=N_1~~~ {\rm otherwise}
    \end{cases}
    \label{eq:example2}
\end{align}
where $N_1\sim {\rm Unif}([-1,1])$ and is independent of $Y$.

\begin{figure}[htbp]
    \centering
    \includegraphics[width=\figSize\columnwidth]{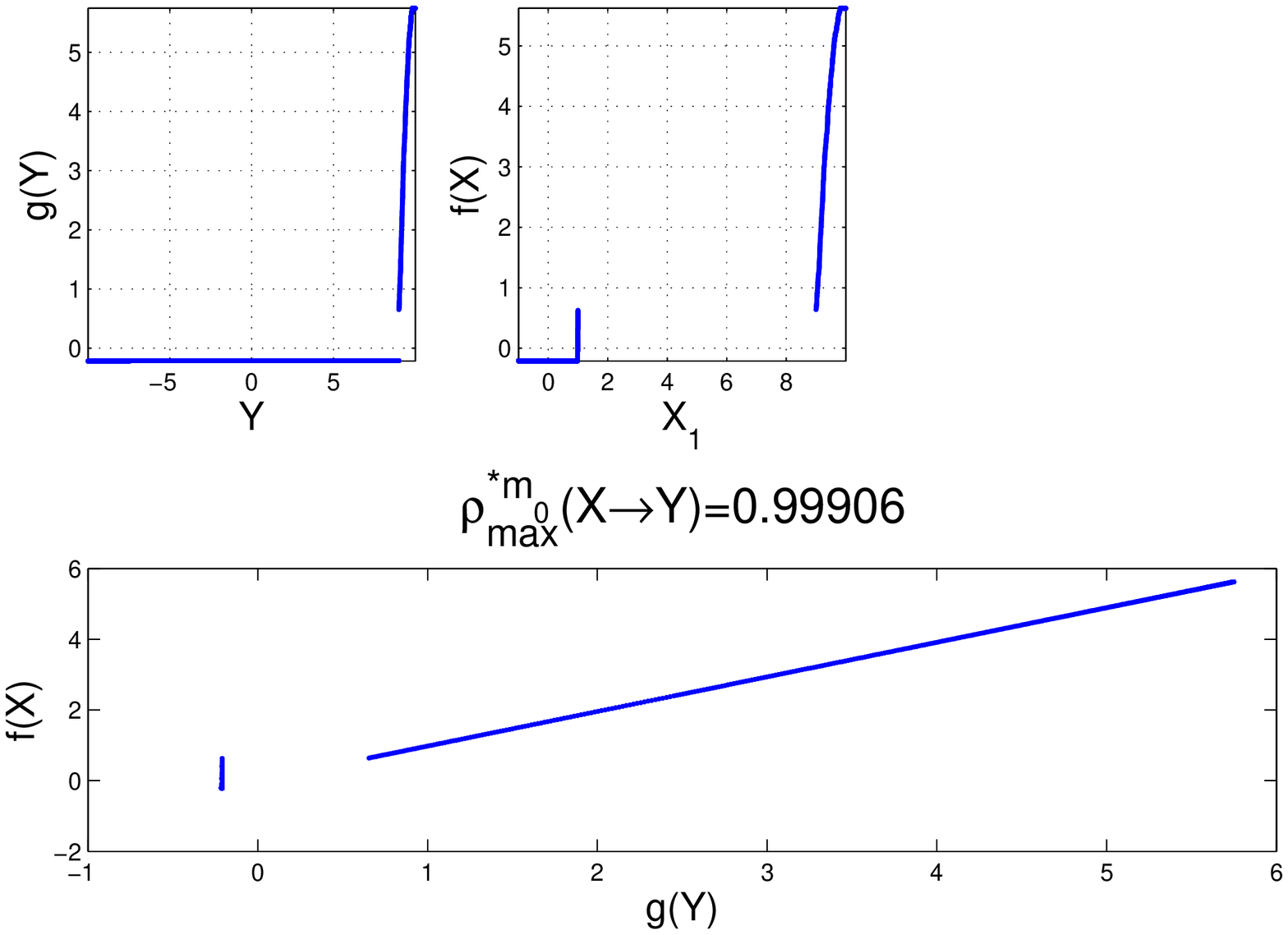}
    \caption{Example 3: Running modified ACE (Algorithm~\ref{Alg:ACE_Semi_0_mono_Single_Predictor}) on $Y$, $X$ with $\kappa=0$.}
    \label{fig:y_x_exmp2_onlyMono}
\end{figure}

Limiting $g$ only to be monotone 
(with no slope limitations)
results in a correlation value of $1$ 
since the optimal solution is to set $g(y)=0$ in the region it cannot be estimated 
and $g(y)=y$ otherwise (and then apply normalization). Clearly, the function $g$ is non-invertible as is depicted in Figure~\ref{fig:y_x_exmp2_onlyMono}.

\begin{figure}[htbp]
    \centering
    \includegraphics[width=\figSize\columnwidth]{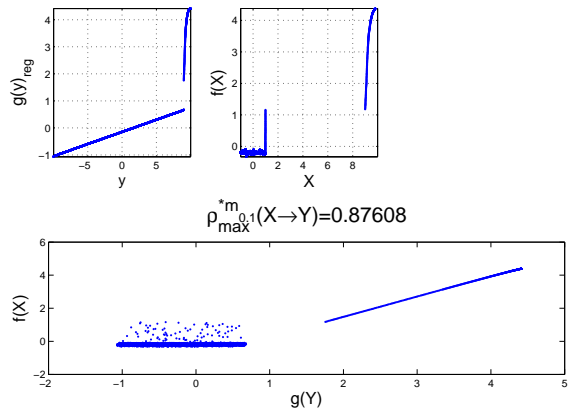}.
    \caption{Example 3: Running regularized ACE (Algorithm~\ref{Alg:ACE_kappa_inc_fixed_norm}) on $Y$, $X$ with $\kappa=0.1$}
    \label{fig:y_x_exmp2_Mono_0_1}
\end{figure}

Next, we ran Algorithm~\ref{Alg:ACE_kappa_inc_fixed_norm} - the regularized ACE algorithm, enforcing a minimal slope of $\kappa=0.1$.
The results are depicted in Figure~\ref{fig:y_x_exmp2_Mono_0_1}.
This example sheds light on the trade-off that exists when setting the value of $\kappa$.
Setting  $\kappa$ to be large
limits the possible gain over the correlation ratio  whereas setting it too low
%
risks overemphasizing regions where the noise is weaker.


\bibliographystyle{IEEEtran}
\bibliography{eladd}


\end{document}

%% file: eli_macros.tex
\newcommand{\var}{\ensuremath{\textrm{var}}}



%% file: rv_defs.tex
\DeclareMathAlphabet{\mathbsf}{OT1}{cmss}{bx}{n}
\DeclareMathAlphabet{\mathssf}{OT1}{cmss}{m}{sl}
\DeclareMathAlphabet{\mathcsf}{OT1}{cmss}{sbc}{n}

\newcommand{\svv}[1]{\mathbf{#1}}

\DeclareSymbolFont{bsfletters}{OT1}{cmss}{bx}{n}  
\DeclareSymbolFont{ssfletters}{OT1}{cmss}{m}{n}
\DeclareMathSymbol{\bsfGamma}{0}{bsfletters}{'000}
\DeclareMathSymbol{\ssfGamma}{0}{ssfletters}{'000}
\DeclareMathSymbol{\bsfDelta}{0}{bsfletters}{'001}
\DeclareMathSymbol{\ssfDelta}{0}{ssfletters}{'001}
\DeclareMathSymbol{\bsfTheta}{0}{bsfletters}{'002}
\DeclareMathSymbol{\ssfTheta}{0}{ssfletters}{'002}
\DeclareMathSymbol{\bsfLambda}{0}{bsfletters}{'003}
\DeclareMathSymbol{\ssfLambda}{0}{ssfletters}{'003}
\DeclareMathSymbol{\bsfXi}{0}{bsfletters}{'004}
\DeclareMathSymbol{\ssfXi}{0}{ssfletters}{'004}
\DeclareMathSymbol{\bsfPi}{0}{bsfletters}{'005}
\DeclareMathSymbol{\ssfPi}{0}{ssfletters}{'005}
\DeclareMathSymbol{\bsfSigma}{0}{bsfletters}{'006}
\DeclareMathSymbol{\ssfSigma}{0}{ssfletters}{'006}
\DeclareMathSymbol{\bsfUpsilon}{0}{bsfletters}{'007}
\DeclareMathSymbol{\ssfUpsilon}{0}{ssfletters}{'007}
\DeclareMathSymbol{\bsfPhi}{0}{bsfletters}{'010}
\DeclareMathSymbol{\ssfPhi}{0}{ssfletters}{'010}
\DeclareMathSymbol{\bsfPsi}{0}{bsfletters}{'011}
\DeclareMathSymbol{\ssfPsi}{0}{ssfletters}{'011}
\DeclareMathSymbol{\bsfOmega}{0}{bsfletters}{'012}
\DeclareMathSymbol{\ssfOmega}{0}{ssfletters}{'012}

%% file: main.bbl
\begin{thebibliography}{10}
\providecommand{\url}[1]{#1}
\csname url@samestyle\endcsname
\providecommand{\newblock}{\relax}
\providecommand{\bibinfo}[2]{#2}
\providecommand{\BIBentrySTDinterwordspacing}{\spaceskip=0pt\relax}
\providecommand{\BIBentryALTinterwordstretchfactor}{4}
\providecommand{\BIBentryALTinterwordspacing}{\spaceskip=\fontdimen2\font plus
\BIBentryALTinterwordstretchfactor\fontdimen3\font minus
  \fontdimen4\font\relax}
\providecommand{\BIBforeignlanguage}[2]{{%
\expandafter\ifx\csname l@#1\endcsname\relax
\typeout{** WARNING: IEEEtran.bst: No hyphenation pattern has been}%
\typeout{** loaded for the language `#1'. Using the pattern for}%
\typeout{** the default language instead.}%
\else
\language=\csname l@#1\endcsname
\fi
#2}}
\providecommand{\BIBdecl}{\relax}
\BIBdecl

\bibitem{cramer2016mathematical}
H.~Cram{\'e}r, \emph{Mathematical methods of statistics (PMS-9)}.\hskip 1em
  plus 0.5em minus 0.4em\relax Princeton University Press, 2016, vol.~9.

\bibitem{renyi1959new}
A.~R{\'e}nyi, ``New version of the probabilistic generalization of the large
  sieve,'' \emph{Acta Mathematica Hungarica}, vol.~10, no. 1-2, pp. 217--226,
  1959.

\bibitem{hirschfeld1935connection}
H.~O. Hirschfeld, ``A connection between correlation and contingency,'' in
  \emph{Mathematical Proceedings of the Cambridge Philosophical Society},
  vol.~31, no.~4.\hskip 1em plus 0.5em minus 0.4em\relax Cambridge University
  Press, 1935, pp. 520--524.

\bibitem{gebelein1941statistische}
H.~Gebelein, ``Das statistische {P}roblem der {K}orrelation als
  {V}ariations-und {E}igenwertproblem und sein {Z}usammenhang mit der
  {A}usgleichsrechnung,'' \emph{ZAMM-Journal of Applied Mathematics and
  Mechanics/Zeitschrift f{\"u}r Angewandte Mathematik und Mechanik}, vol.~21,
  no.~6, pp. 364--379, 1941.

\bibitem{renyi1959measures}
A.~R{\'e}nyi, ``On measures of dependence,'' \emph{Acta Mathematica Hungarica},
  vol.~10, no. 3-4, pp. 441--451, 1959.

\bibitem{breiman1985estimating}
L.~Breiman and J.~H. Friedman, ``Estimating optimal transformations for
  multiple regression and correlation,'' \emph{Journal of the American
  Statistical Association}, vol.~80, no. 391, pp. 580--598, 1985.

\bibitem{samuel2001correlation}
D.~Drouet-Mari and S.~Kotz, \emph{Correlation and dependence}.\hskip 1em plus
  0.5em minus 0.4em\relax World Scientific, 2001.

\bibitem{hall1967characterizing}
W.~Hall, \emph{On characterizing dependence in joint distributions}.\hskip 1em
  plus 0.5em minus 0.4em\relax University of North Carolina, Department of
  Statistics, 1967.

\bibitem{kimeldorf1978monotone}
G.~Kimeldorf and A.~R. Sampson, ``Monotone dependence,'' \emph{The Annals of
  Statistics}, pp. 895--903, 1978.

\bibitem{ramsay1988monotone}
J.~O. Ramsay, ``Monotone regression splines in action,'' \emph{Statistical
  Science}, vol.~3, no.~4, pp. 425--441, 1988.

\bibitem{hastie1988monotone}
T.~Hastie and R.~Tibshirani, ``[monotone regression splines in action]:
  Comment,'' \emph{Statistical Science}, vol.~3, no.~4, pp. 450--456, 1988.

\bibitem{tebbe1968uncertainty}
D.~Tebbe and S.~Dwyer, ``Uncertainty and the probability of error (corresp.),''
  \emph{IEEE Transactions on Information Theory}, vol.~14, no.~3, pp. 516--518,
  1968.

\bibitem{breiman1988monotone}
L.~Breiman, ``[monotone regression splines in action]: Comment,''
  \emph{Statistical Science}, vol.~3, no.~4, pp. 442--445, 1988.

\bibitem{li2016true}
H.~Li, ``A true measure of dependence,'' University Library of Munich, Germany,
  Tech. Rep., 2016.

\bibitem{lancaster1957some}
H.~O. Lancaster, ``Some properties of the bivariate normal distribution
  considered in the form of a contingency table,'' \emph{Biometrika}, vol.~44,
  no. 1/2, pp. 289--292, 1957.

\bibitem{yu2008maximal}
Y.~Yu, ``On the maximal correlation coefficient,'' \emph{Statistics \&
  Probability Letters}, vol.~78, no.~9, pp. 1072--1075, 2008.

\bibitem{schweizer1981nonparametric}
B.~Schweizer, E.~F. Wolff \emph{et~al.}, ``On nonparametric measures of
  dependence for random variables,'' \emph{The annals of statistics}, vol.~9,
  no.~4, pp. 879--885, 1981.

\bibitem{voss1997reconstruction}
H.~Voss and J.~Kurths, ``Reconstruction of non-linear time delay models from
  data by the use of optimal transformations,'' \emph{Physics Letters A}, vol.
  234, no.~5, pp. 336--344, 1997.

\end{thebibliography}
